\documentclass[twocolumn,showpacs,preprintnumbers,amsmath,amssymb]{revtex4}
\usepackage{graphicx} 
\usepackage{dcolumn} 
\usepackage{bm} 
\usepackage{amsmath,amssymb}

\begin{document}

\title{Intrinsic thermal vibrations of suspended doubly clamped single-wall carbon nanotubes}

\author{B.\ Babi{\'c}}
\author{J.\ Furer}
\author{S.\ Sahoo}
\author{Sh.\ Farhangfar}
\author{C. Sch{\"o}nenberger}
\email{Christian.Schoenenberger@unibas.ch}
\homepage{www.unibas.ch/phys-meso}
\affiliation{Institut f\"ur Physik, Universit\"at Basel, Klingelbergstr.~82, CH-4056 Basel, Switzerland }
\date{\today}

\begin{abstract}
We report the observation of thermally driven mechanical vibrations of
suspended doubly clamped carbon nanotubes, grown by chemical vapor deposition (CVD). Several
experimental procedures are used to suspend carbon nanotubes. The
vibration is observed as a blurring in images taken with a scanning electron microscope.
The measured vibration amplitudes are compared with a model based on linear continuum mechanics.
\end{abstract}

\maketitle

Carbon nanotubes (NTs) form a material with unique mechanical
properties \cite{Ebbesen,Poncharal,Wagner,Ruoff,Walters}. The high
Young's modulus and low specific weight
qualify single-wall carbon nanotubes (SWNTs) as
ultimate mechanical resonators.
Similar to lithographically patterned \mbox{SiC} beams, whose
resonance frequency has recently crossed the border from MHz to
GHz \cite{Roukes1}, it would be highly desirable to integrate NTs
into nanoelectromechanical systems (NEMSs) and to electrically
excite the mechanical vibration modes \cite{Roukes2}.
A first step in this direction has been the observation of electrically driven
mechanical vibrations of multi-wall carbon nanotubes~\cite{Poncharal}.
Nanometer-sized resonators oscillate at high frequencies, but
simultaneously have small vibration amplitudes, which are
difficult to measure. At cryogenic temperatures, the resonant
adsorption of an external electromagnetic field could successfully
be measured using superconducting elements attached to a freely
suspended NT \cite{Reulet}. At room temperature a tunnelling probe
in the form of, for example, an STM tip would be a
versatile detector. Integrating a sensitive measuring transducer
with a NT nanomechanical oscillator is however challenging. In a first
step, it would be desirable if the mechanical vibrations could be
imaged directly. Here, we report on the observation of thermal
vibrations of suspended doubly clamped SWNTs, imaged by scanning
electron microscopy (SEM).

Thermally driven excitations of multi-wall carbon nanotubes (MWNTs), clamped at one end only,
were first investigated by Treacy \textit{et al.} \cite{Ebbesen}.
The mechanical oscillation appeared in the images, which were collected with a transmission
electron microscope (TEM), as a blurring that increased towards the free end of the MWNTs.

\begin{figure}
\begin{center}
\end{center}
\caption{\label{Figure 1.} Schematic drawings of doubly clamped vibrating
SWNT which are suspended by different methods. (a) top view, (b-d) side views.}
\end{figure}

In order to see whether a similar experiment is possible with doubly clamped SWNTs,
we will first estimate the expected amplitude in thermal equilibrium at room temperature.
A schematics with coordinate system is shown in Fig.~1a.
Assuming that linear continuum mechanics is a good approximation, the equation of
motion for the vertical displacement $\xi$ is given by \cite{Liftshiz}
\begin{equation}\label{eq:diff-equation}
\frac{\partial^2\xi}{\partial t^2}+ \left( \frac{YI}{\rho A}\right)
\frac{\partial^4\xi}{\partial x^4}=0 \text{.}
\end{equation}
Here, $\rho$ is the mass density, $A$ the cross-sectional area, $Y$ the
Young's modulus, and $I=\pi d^4/64$ the moment of inertia, which depends only on the
diameter $d$. Applying the proper boundary conditions, the spectrum of eigenfrequencies is obtained:
\begin{equation}\label{eq:eigenfrequency}
\omega_i=\frac{\beta_i^2}{L^2}\sqrt{\frac{YI}{\pi d\rho_{2d}}}\text{~~~}(i=1,2,3\dots)\text{~,}
\end{equation}
where $L$ is the suspended length, $\rho_{2d}$ is the surface mass density
of a graphite sheet \mbox{($7.7\cdot 10^{-7}$\,kgm$^{-2}$)}and $\beta_{1}=4.73$, $\beta_{2}=7.85$, and
$\beta_{3}=11.0$ for the first three modes.

The equipartition theorem predicts that each vibration mode carries the energy
$k_BT$ in thermal equilibrium at temperature $T$, where $k_B$ is the Boltzmann constant.
Together with the appropriate solutions of Eq.~(\ref{eq:diff-equation}), one obtains an
expression for the variance of the maximum deflection amplitude,
which for the fundamental frequency \mbox{($i=1$)} occurs in the middle:
\begin{equation}\label{eq:variance}
\sigma_1^{2} \equiv <\xi_1^{2}(L/2)>=\frac{k_B T L^3}{\gamma_1 YI},
\end{equation}
where $\gamma_1=192$ \cite{Blanter}. Table~\ref{table1} summarizes
the eigenfrequencies and the thermal vibration amplitudes at room
temperature of a `typical' SWNT with diameter \mbox{$d=1.5$\,nm}
and Young's modulus \mbox{$Y=1$\,TPa} for different (practically
feasible) suspension lengths \mbox{$L=0.2-5$\,$\mu$m}. This table
demonstrates that thermal vibration amplitudes can be of appreciable
magnitude, of order \mbox{$\approx 10$\,nm} \mbox{($L=1$\,$\mu$m)}. Since
state-of-the-art scanning electron microscopes have resolutions
well below \mbox{$10$\,nm}, thermal
vibration should appear on SEM images.
\begin{center}
\begin{table}[h]
\caption{Characteristic quantities of suspended SWNT
(\mbox{$d=1.5$\,nm}, \mbox{$Y=1$\,TPa}). The eigenfrequencies and maximum
thermal amplitudes are calculated using relations
(\ref{eq:eigenfrequency}) and (\ref{eq:variance}), respectively.}
\begin{tabular}{ccc}\label{table1}
  \mbox{\textit{L}\,($\mu$m)} &~~\mbox{$\omega$\,(MHz)}   & ~~ \mbox{$\sigma_1$\,(nm)} \\
\hline\hline
  0.2 & 4600 & 0.8 \\
  0.5 & 740 & 3.4 \\
  1 & 185 & 9.3  \\
  3 & 21 & 48.3  \\
  5 & 7.4 & 104 \\
  \hline
\end{tabular}
\end{table}
\end{center}
There are already many reports on the fabrication of suspended NTs. For example,
SWNTs were grown between distant silicon towers \cite{Dai1,
Dai3}, spread over metal posts \cite{Roth}, or grown over solid
terraces \cite{Dai2} and etched trenches \cite{Walters}. Though
devices with suspension lengths of \mbox{$L \gtrsim
5$\,$\mu$m} were realized and imaged with SEM, the thermal
vibration has surprisingly not yet been reported, although it
should readily have shown up in respective SEM images, provided
the reported SWNTs were {\em single} SWNTs. In the
work of Dai and coworkers \cite{Dai1,Dai2,Dai3}, the SWNTs were
coated with a metal layer to increase the contrast in the SEM,
whereas others have explicitly reported on suspended {\em ropes}
of SWNTs or MWNT \cite{Ruoff,Roth}, which are inherently stiffer.

Carbon nanotubes are synthesized by chemical vapor deposition
(CVD) as previously reported \cite{us}. We would like to emphasize
that not all grown NTs are individual SWNTs. This will be
explained further in the text. To account for the possible
influence of substrate during imaging in SEM, we have suspended
NTs using three different methods.

Method {\rm I}, shown in Fig.~1b, is based on the
work of Nyg\aa rd \textit{et al.} \cite{Nygard}. The NTs are grown on
thermally oxidized \mbox{($400$\,nm)} Si substrates. Electrical
contacts are patterned by electron-beam lithography (EBL),
followed by evaporation (Ti/Au) and lift-off.
The \mbox{SiO$_2$} is etched in buffered HF \cite{us1}. To stop etching, the sample
is heavily rinsed in water followed by isopropanol. With this
method we find it possible to suspend NTs over distances up to
\mbox{$1$\,$\mu$m}. For larger lengths, the surface
tension of the etchant tends to pull the NT down to the substrate.

In method {\rm II}, shown in Fig.~1c, the NTs
are grown across predefined trenches. We start with a Si substrate
with layers of \mbox{$800$\,nm} of SiO$_2$ and \mbox{$200$\,nm} of
\mbox{Si$_3$N$_4$}. Slits of width \mbox{$1-5$\,$\mu$m} and length
\mbox{$10$\,$\mu$m} are first etched into the top
\mbox{Si$_3$N$_4$} layer using a \mbox{CHF$_3$-based} plasma
etching process \cite{Hoss}. Next, the
slit is further wet-etched into SiO$_2$ and the Si substrate
using HF and KOH \cite{us2}, respectively.
This results in deep trenches \mbox{$\sim 3.5$\,$\mu$m}, a
prerequisite for NTs to bridge the trenches in the CVD growth
process.

In method {\rm III}, shown in Fig.~1d,
slits are defined in \mbox{Si$_3$N$_4$} membranes of thickness \mbox{$150$\,nm}
and lateral size \mbox{$0.5$\,mm} following a similar procedure as in method~{\rm II}.

The key difference between the three methods is the depth of suspension. It is
\mbox{$400$\,nm}, \mbox{$3.5$\,$\mu$m}, and \mbox{$\infty$} for methods {\rm I-III}, respectively.
The samples are imaged with SEM (Philips XL30 FEG) at room temperature.
To generate an image, a focused electron beam is raster scanned.

To deduce the vibration amplitude quantitatively two assumptions
have to be made: 1) the intensity profile of the electron beam
centered at coordinate \mbox{$(x,y)$} has a Gaussian distribution
and 2) the measured intensity of secondary electrons reflects the (time-averaged)
probability $P(x,y)\equiv P_x(\xi)$ to find the NT at position
\mbox{$(x,y)$} convoluted with the intensity profile of
the primary beam. 1) is a convenient assumption and 2) should
hold, because scanning in SEM is slow as compared to the vibration
of the NT. The latter results in a blurring of the NT in SEM
images. An example of a vibrating suspended NT is shown in
Fig.~2a. The vibration is observed as a blurring, which is largest
in the middle. In contrast, the NT appears sharp at the edges of
the trench, limited by the finite resolution of the SEM. To deduce
the vibration amplitude, more precisely the variance
\mbox{$\sigma^2(x)\equiv\langle\xi^2(x)\rangle$}, we note that
$P_x(\xi)$ is Gaussian and determined by Boltzmann statistics. The
deconvolution is simple because of assumption 1). We only
need to extract $\sigma^2(x)$ from the intensity distribution of
the SEM image perpendicular to the NT and subtract $\sigma^2(0)$.
To do so, we average the intensity profile in $\Delta x$ slices as
shown in Fig.~2b and fit it to a Gaussian. Such an analysis was
first done for MWNT cantilevers by Krishnan {\it et al.}
\cite{Ebbesen1}.

\begin{figure}
\begin{center}
\end{center}
\caption{\label{Figure 2.} (a) SEM image of a vibrating SWNT grown
over a trench. A strong blurring is clearly visible (indicated
by arrows), which is a consequence of intrinsic thermal vibrations.
(b) Another vibrating NT, whose root-mean-square displacement along $x$
is plotted in (c). Circles are measured points and the curves
represent fits.}
\end{figure}

Figure 2b shows a SEM image of a suspended doubly clamped
vibrating NT fabricated by method~{\rm I}. The free suspension
length is relatively short, i.e. \mbox{$L\approx 650$\,nm}.
Applying the analysis procedure mentioned above, the maximum rms
vibration amplitude is determined to be \mbox{$\sigma=27\pm
5$\,nm}. We have also analyzed $\sigma$ as a function of $x$ and
compare the result with analytical curves for the first three
eigenmodes in Fig.~2c. The agreement between the measured
points and the theoretical curves is reasonably good. Matching
between experiment and theory is improved if the first {\em and}
second modes are taken into account, each of which carries $kT$
energy. Contributions from higher order modes decay very rapidly
and can be neglected. Note, there is one fitting parameter $Yd^4$,
which will be discussed below.

\begin{figure}
\begin{center}
\end{center}
\caption{\label{Figure. 3.} (a) SEM image of long
(\mbox{\textit{L}\,$\approx6.2$\,nm}) vibrating SWNT grown over a slit in
a \mbox{Si$_3$N$_4$} membrane. In (b) three NTs are imaged simultaneously.
Only the middle one is vibrating. A white circle indicates branching of the lower NT into two NTs.}
\end{figure}

Figure 3a shows another NT grown over \mbox{Si$_3$N$_4$} membrane.
Here, the suspension length is rather large, i.e. \mbox{$L\approx
6.2$\,$\mu$m}. Correspondingly, the observed blurring is much
larger. The maximum rms vibration amounts to \mbox{$\sigma=80 \pm
5$\,nm}. Applying Eq.~(\ref{eq:variance}) and assuming the typical
high Young's modulus value of SWNTs of \mbox{\textit{Y}=1\,TPa}
the diameter of this NT is estimated to be \mbox{$d=2 \pm
0.5$\,nm}.

The SEM image displayed in Fig.~3b shows three suspended NTs.
Though grown in one run, only one NT seems to vibrate, namely the
middle one. This, at first sight surprising result, points to a
variability of NTs that are grown during one and the same process.
The only parameter in our experiment, which is not predetermined,
is $Yd^4$, see Eq.~(\ref{eq:variance}). Though different values
for the Young's modulus were reported, we suspect that the
diameter $d$ is the cause for the variability, because it enters
in the fourth power. The absence of visible vibrations for the
upper and lower NT in Fig.~3b suggests that these have a larger
diameter. They may be multi-wall nanotubes or ropes of tubes. In
fact, the lower one must be a rope, because a clear branching is
observed at the right end (highlighted by a circle). Having looked
through a large number of samples, the fraction of vibrating tubes
is very small \mbox{(a few \%)}. This is a clear indication that
not all of the grown NTs are SWNTs.

\begin{center}
\begin{table}[h]
 \caption{Properties of some vibrating NTs. $L$ is
the suspended length, $\sigma$ the measured maximum rms vibration,
$Yd^4$ obtained using Eq.~\ref{eq:variance}, $Y_{1.6}$ Youngs'
modulus assuming \mbox{$d=1.6$\,nm} (see text), and $d_{1}$ the NT
diameter assuming \mbox{$Y=1$\,TPa}.} \label{table2}
\begin{tabular}{cccccc}
$L$ & $\sigma$ &$Yd^4$ & $Y_{1.6}$ & $d_{1}$ &  Method\\
($\mu$m) & (nm)& (GPa(nm)$^4$)& (GPa) & (nm) & \\
\hline\hline
  0.55 & 25 & 117 & 18 & 0.58 & I \\
  0.63 & 27 & 150 & 23 & 0.62 & I\\
  1.35 & 16 & 4221 & 644 & 1.4 & III \\
  4.05 & 85 & 4038 & 616 & 1.4 & II \\
  4.30 & 90 & 4311 & 658 & 1.45 & II \\
  6.25 & 80 & 16754 & 2556 & 2.0 & III \\\hline\\
\end{tabular}
\end{table}
\end{center}

We summarize the measured rms vibration of several NTs in
table~\ref{table2}. Determined are $L$ and $\sigma(L/2)$. Using
Eq.~(\ref{eq:variance}), we
obtain an estimate for $Yd^4$, which is given in the third column.
What is immediately noticed is the large spread in $Yd^4$
of more than two orders of magnitude.
Unfortunately, we are not able to unambiguously deduce the Young's
modulus $Y$ and diameter $d$, independently. We have tried to
measure the diameter using atomic-force microscopy (AFM). Due to
surface roughness and the strong $d^4$ dependence, the error bar
is too large to deduce $Y$ with an acceptable accuracy. For the
discussion we instead rely on an average diameter for SWNTs, which
we have obtained from electrical measurements of contacted
semiconducting NTs \cite{us}. We have analyzed the band-gap, which
is inversely proportional to the diameter $d$,
of more than $10$ semiconducting
SWNTs and obtained as an average \mbox{$d=1.6 \pm 0.3$\,nm}. We
note, that taking this diameter, the estimated Young's modulus
(column $4$ in table~\ref{table2}, denoted by $Y_{1.6}$) has an
accuracy of `only' \mbox{$75$\,\%}. Well graphitized NTs have a
large Young's modulus. For example, \mbox{$Y = 1.4 \pm 0.4$\,TPa}
was reported for SWNTs grown by laser ablation \cite{Ebbesen1},
whereas \mbox{$1$\,TPa} was found in simulations independent of
helicity and number of shells \cite{Lu}. In column $5$ of
table~\ref{table2} we therefore also list the diameter $d_1$,
which we deduce from the measured $Yd^4$, assuming {$Y=1$\,TPa}.
$d_1$ is varying between \mbox{0.58} to
\mbox{$2.0$\,nm}. Since we have never observed SWNTs with
diameters \mbox{$< 1$\,nm} in TEM, the first two NTs (row $1$ and
$2$), both belonging to samples prepared by method {\rm I}, cannot
have a large Young's modulus \mbox{$Y\sim 1$\,TPa}. Taking $d$ to
be \mbox{$1.6$\,nm} leads to a modulus of only
\mbox{$Y_{1.6}\approx 20$\,GPa}. Because method~{\rm I} uses
HF-etching we suspect that the NT's are affected during this
process step. It is possible that CVD-grown SWNTs are not perfect
so that wet etching can proceed starting at defect sites. In
contrast to method~{\rm I}, the as-grown CVD NTs of methods~{\rm
II} and {\rm III} yield consistent results, which are in agreement with a
large Young's modulus of \mbox{$1$\,TPa} and with the diameter,
which we have deduced by electrical measurements. Though we
observe ropes and small diameter MWNTs (only a few number of
shells) in TEM, their diameter is typically larger than
\mbox{$2$\,nm}. This strongly suggests that the NTs of row $3-6$
in table~\ref{table2} are {\em single}-wall carbon nanotubes.

To our knowledge there are no reports on the Young's modulus of
CVD-grown SWNT. Though we are not able to accurately determine
$Y$, our results suggest that CVD-grown SWNTs can have a large
modulus of order \mbox{$Y\approx 1$\,TPa}. The exception are
wet-etched NTs, for which our data suggest \mbox{$Y\ll 1$\,TPa}.
Small Young's modulus have previously been reported for CVD-grown
MWNTs \cite{Salvetat}.

In conclusion, we have demonstrated that it is possible to observe
thermally driven vibrations of suspended doubly clamped SWNTs in
SEM. From the measured rms vibration amplitude, the Young's
modulus $Y$ of CVD-grown SWNTs has been estimated. Only a small
fraction of suspended NTs are seen to vibrate, although
they are suspended over a comparable length and grown at the same
time. This suggests that the majority of grown tubes are {\em not}
single SWNTs, but rather ropes and MWNTs, a finding, which is supported
by TEM. We suspect that this is the reason why thermal vibrations
of SWNTs has not already been observed before.

\begin{acknowledgments}
We acknowledge support and technical assistance with TEM imaging
by A. Engel, G. Gantenbein and H. Stahlberg of the Biocenter
and discussions with L.
Forr{\'o}. This work has been supported by COST (BBW), the Swiss
NFS and the NCCR on Nanoscience.
\end{acknowledgments}



\begin{thebibliography}{99}
\bibitem{us1}
Buffered hydrofluoric acid is made according to the following recepie:
\mbox{$28$\,ml} HF + \mbox{$107$\,ml} H$_2$0 + \mbox{$113$\,g} NH$_4$F.
The etch rate for \mbox{SiO$_2$} is \mbox{$50$\,nm/min}.
\bibitem{us2}
The etch ratio of \mbox{Si} to \mbox{SiO$_2$} is \mbox{$\approx
500$}, for \mbox{$20$\,\%} weight mass of KOH at
\mbox{$60$\,$^0$C}.
\end{thebibliography}
\end{document}